## Highlights

- Increases in GHG concentrations will lead to more aridity over the IP.
- Drying trend is expected to behave differently along the year and across the IP.
- Soil drying might be mostly driven by the reduction in large-scale precipitation.
- In northern IP, spring evapotranspiration would amplify the soil drying.
- Important changes will occur due to GHG rising trends at high altitude during winter.

# Future Changes in Land and Atmospheric Variables: An Analysis of their Couplings in the Iberian Peninsula

Matilde García-Valdecasas Ojeda, Patricio Yeste, Sonia Raquel Gámiz-Fortis, Yolanda Castro-Díez, and María Jesús Esteban-Parra

Department of Applied Physics. University of Granada. mgvaldecasas@ugr.es

## ABSTRACT

This work investigates climate-change projections over a transitional region between dry and wet climates, the Iberian Peninsula (IP). With this purpose, the Weather Research and Forecasting (WRF) model, driven by two global climate models (CCSM4 and MPI-ESM-LR) previously bias-corrected, was used to generate high-resolution climate information. Simulations were carried out for two periods, 1980–2014 and 2071–2100, and under two representative concentration pathways (RCP4.5 and RCP8.5). The analysis focused on changes in land-surface processes, their causes, and the potential impact on the climate system. To achieve this, seasonal projected changes of land-surface (soil moisture and surface evapotranspiration) and atmospheric variables involved in the hydrologic (i.e., precipitation and runoff) and energy balance (i.e., temperature and solar incoming radiation) were investigated. The results reveal that the IP is likely to experience a soil dryness by the end of the 21$^{st}$ century, particularly during summer and fall, more apparent in the southern IP, and stronger under the RCP8.5. However, such trends would have different implications throughout the year and directly affect the surface evapotranspiration. Moreover, soil-drying trends are mainly associated with reductions in the large-scale precipitation during spring, summer, and fall and by enhanced evapotranspiration particularly in spring over the northwestern IP. In addition, the results show notably changes in soil conditions at high altitude, particularly during winter, which may alter the land-atmosphere processes that currently occur in these regions. In this context, noteworthy changes in the climate system are expected, leading to adverse impacts on water resources and temperature. The results highlight the complex and nonlinear nature of land-atmosphere interactions in regions such as the IP, which is a tremendous challenge for adequately developing mitigation and adaptation strategies to anthropogenic climate change.

## 1. Introduction

The rising trend of the temperature caused by anthropogenic greenhouse gas (GHG) emissions is expected to cause important changes to the global water cycle (Sheffield and Wood 2008). However, substantial uncertainties persist concerning the magnitude of the effects on the different hydroclimatic variables, particularly in mid-latitude regions (Greve et al., 2018). Hence, evaluating changes in water availability is a great challenge for the proper development of water management strategies.

There is a strong consensus on the high relevance of the land surface state corresponding to the regional and local climate (Berg et al., 2016; Jaeger and Seneviratne, 2011; Menéndez et al., 2019; among others).In this regard, soil moisture is an essential factor that may alter both atmospheric and land variables, and its influence has been noted for long periods (Khodayar et al., 2015) and over large areas (Zampieri et al., 2009). This is especially true in transitional regions between wet and dry climates, where soil moisture controls the changes in the partitioning of radiative energy into sensible and latent heat fluxes, leading to land-atmosphere feedbacks. In these regions, negative anomalies of soil moisture may exacerbate extreme events, such as drought (Quesada et al., 2012) and heatwaves (Miralles et al., 2014). Hence, land water storage largely implicates the resulting surface climate, altering the temperature (Vogel et al., 2017), the boundary layer stability (Dirmeyer et al., 2013), and the subsequent precipitation (Guo et al., 2006).

Recent studies have addressed the analysis of the physical mechanisms through which the enhanced GHG concentrations will influence the climate system in the future, and the major relevance of the soil moisture on the future climate is well-known. For instance, Orth and Seneviratne (2017) revealed that soil moisture variability will affect the climate similarly to that by the sea surface temperature variability over mid-latitude regions, with a stronger influence in terms of extreme temperature and precipitation. Diffenbaugh et al. (2005) evidenced that soil dryness amplifies the effects of the enhanced GHG concentrations on the extreme temperature and precipitation over the contiguous United States via land-atmosphere feedbacks. For Europe, Seneviratne et al. (2006) pointed out that changes in temperature variability will be at least partly due to the increase in the strength of the land-atmosphere coupling. Although this latter study analyzed the soil moisture effects on the climate over the European region, the authors emphasized the results found over regions where energy-limited regimes appeared over their current simulations (eastern and central Europe). Other studies have further investigated the impact of soil conditions on climate variability using prescribed soil moisture, finding that this variable is likely to modify both the hydrologic and the energy balance (Jaeger and Seneviratne, 2011; Seneviratne et al., 2013; Vogel et al., 2018; among others), leading to important changes in the mean and extreme values of temperature and precipitation. For the Iberian Peninsula (IP), Jerez et al. (2012) studied the effect of using different land-surface models (LSMs) on regional climate projections. They evidenced that the land-surface processes are crucial to adequately project both the mean and variability of temperature, precipitation, and wind. Given that the soil moisture influence is more noticeable over dry seasons, most of these studies focused on boreal summer (June–August). However, in a recent work, Ruosteenoja et al. (2018) recognized the importance of studying drying trends throughout the year as soil depletion will have different implications depending on the season.

Global climate models (GCMs) constitute the basis to predict changes in the climate in the context of ongoing global warming as a noteworthy tool to reproduce the large-scale circulation (Giorgi et al., 2011). However, the capability of GCMs in reproducing regional climates remains inadequate due to their coarse resolution. In this regard, regional climate models (RCMs) have proved to add value in simulating the regional climate; thus, are more appropriate for the generation of climate-change projections at a higher spatial resolution (van der Linden et al., 2019), and hence, for detecting local feedbacks.

In this study, we analyzed the projected changes in soil conditions by the end of this century under two different representative concentration pathways (RCPs) to elucidate how these changes would influence atmospheric conditions in a context of a transitional climate. To achieve this, we conducted a set of high-resolution projections using the Weather Research and Forecasting (WRF) model. Simulating the climate at high resolution is particularly relevant in regions such as the IP, where a high spatiotemporal variability in precipitation occurs due to different factors. More specifically, the IP is a topographically complex region with extensive coasts, located between two climate regions (the subtropical and the mid-latitude areas) as well as between two completely different water masses (the Mediterranean Sea and the Atlantic Ocean). Moreover, the study of changes in soil conditions are of major relevance as the IP has been considered a hotspot regarding its current soil state conditions; thus, is particularly vulnerable to the increase of GHGs. Moreover, studies such as that by Zampieri et al. (2009) evidenced the link between spring/early summer soil conditions over the Mediterranean region with the development of heatwaves over continental Europe. Therefore, the study of future soil conditions across the IP is of high interest for the entire European region.

## 2. Data and methods

### 2.1. Weather research and forecasting setup

The WRF-ARW model (Skamarock et al., 2008) version 3.6.1 was selected to complete a set of high-resolution climate projections. All WRF simulations were carried out using two “one-way” nested domains (Fig. 1): the outer domain (d01) corresponds to a domain with 126 × 123 grid-points that covers the EURO-CORDEX region (Jacob et al., 2014) at 0.44° of spatial resolution. The finer domain (d02), configured with 221 × 221 grid-points, is centered over the IP at 0.088° of spatial resolution, and both domains were configured using 41 vertical levels with the top of the atmosphere set to 10 hPa. A spectral nudging approach was used by adjusting waves above 600 km (Messmer et al., 2017). This was applied only for the coarser domain, and above the planetary boundary layer (PBL); thus, allowing the RCM to perform its own internal dynamic in the finer domain (Argüeso et al., 2011).

The simulations were performed using two different GCMs from the Coupled Model Intercomparison Project phase 5 (CMIP5) as the lateral boundary conditions, which were previously bias-corrected, the NCAR's CCSM4 (Gent et al., 2011), and the Max Plank Institute MPI-ESM-LR (Giorgetta et al., 2013) in its run r1i1p1. Bias-corrected outputs from the NCAR's CCSM4 (Monaghan et al., 2014), which follow the approach by Bruyère et al. (2014), are available in the format required to run the WRF. In the same way, we also corrected the outputs from the MPI-ESM-LR model using this same method.

To analyze future projections over the IP, we selected the period from 2071 to 2100 using two RCPs. On the one hand, RCP4.5 was used as a stabilization scenario that considers an increase of GHG concentrations corresponding to a global temperature increase of approximately 1.8 °C by the end of this century. On the other hand, RCP8.5 was used as it is the most severe emission scenario, with a continuous increase of GHGs and a global temperature increase of approximately 4 °C by 2100 (Moss et al., 2010). Additionally, simulations of the current state were generated to quantify future changes in relation to the present using the period from 1980 to 2014. To complete these simulations, the outputs from RCP8.5 were used from 2006 to 2014. This RCP adequately describes the actual present conditions, as reported by Granier et al. (2011).

Several authors (Argüeso et al., 2011; Jerez et al., 2013; Kotlarski et al. 2014) have identified the major role played by the set of parameterizations to better represent the climate in a particular region. This is particularly true over complex domains such as the IP. For this reason, a thorough sensitivity study to investigate the best combination of parameterizations for simulating the climate in the IP was already carried out (García-Valdecasas Ojeda et al., 2015). The selected parameterization set has also been successfully used for the representation of spatiotemporal patterns of droughts over the Spanish region (García-Valdecasas Ojeda et al., 2017), which are strongly related to land-surface processes (Quesada et al., 2012). Moreover, the parameterization combinations chosen here agree with previous studies performed over the same region (Argüeso et al., 2011, 2012a, 2012b), namely: the Betts-Miller-Janjic (Betts and Miller, 1986; Janjić, 1994) for convection, Convective Asymmetric Model version 2 (Pleim, 2007) for PBL, and the WRF single-moment-three-class schemes (Hong et al., 2004) for microphysics. The long- and shortwave radiations were parameterized using the Community Atmosphere Model 3.0 (Collins et al., 2004).

Soil processes have been modeled by the Noah LSM (Chen and Dudhia 2001), which computes water and energy fluxes using four different layers (0–10, 10–40, 40–100, and 100–200 cm). Land and atmosphere are coupled through the water balance equation, the surface layer stability, and the surface energy balance (Greve et al., 2013). The Noah LSM uses three key inputs to correctly determine soil processes (vegetation type, texture, and slope), through which different soil parameters (e.g., albedo, leaf area index or canopy resistance) are added by lookup tables. In this study, the soil textures from the Food and Agriculture Organization (FAO) soil datasets (Miller and White, 1998) and the MODIS Land Cover from the International Geosphere-Biosphere Programme at a resolution of 30″ were used (Fig. 1S).

The ability to characterize the atmospheric and soil-related variables from the simulations in this study was extensively analyzed in García-Valdecasas Ojeda (2018) and García-Valdecasas Ojeda et al., (2020). One of their main conclusions was that the simulations obtained from WRF forced by both GCMs reproduced the main spatiotemporal patterns of the variables analyzed over the IP acceptably well, which can then be used to project the climate over the IP.

**2.2. Soil-state variables**

In this work, changes in two soil-related variables, the soil moisture and the surface evapotranspiration (SFCEVP), were examined. To study the soil wetness, the soil moisture index (SMI) at the upper 1 m of soil, which is the most hydrologically active soil region (Giorgi and Mearns, 1999), was computed. The SMI was selected instead of the soil moisture to better represent the total soil water available to plants, thus, making the comparison with the evapotranspiration easier. As defined in Seneviratne et al. (2010), the SMI can be understood as:

$$SMI = \frac{\theta - \theta_{wp}}{\theta_{fc} - \theta wp} \quad (1)$$

where $\theta$ , $\theta_{wp}$ , and $\theta_{fc}$ are the modeled volumetric soil moisture, the volumetric soil water content at the wilting point, and the volumetric soil moisture at field capacity, respectively.

The analysis focused on a direct grid-to-grid comparison of the long-term mean values of the selected soil-related variables at seasonal scale, which is adequate to study the potential soil drying (Ruosteenoja et al., 2018). As our interest was related to high-resolution products, the analysis was performed only with the WRF outputs from the inner domain (d02).

The projected changes were examined using the delta-change approach (Hay et al., 2000) through the differences between the future and the current periods. To determine the significance of these differences, the non-parametric Wilcoxon-Mann-Whitney Rank Sum test was applied (Wilks, 2006) at the 90% confidence level. This test considers the null hypothesis that the future and present time series come from continuous distributions with equal medians.

**2.3. Land-atmosphere coupling diagnostic**

Due to the high spatiotemporal variability that occurs in the IP in terms of land-surface processes, we also investigated how these GCM-driven simulations capture land-atmosphere coupling. For this purpose, interannual temporal correlations between the seasonally averaged SMI and the SFCEVP were used as a coupling metric (Dirmeyer et al., 2013; Diro and Sushama et al., 2017). Thus, a negative SMI-SFCEVP correlation indicates that the soil moisture content is enough to satisfy the water demand by the atmosphere. Therefore, atmospheric conditions control the soil moisture content, and an increase (decrease) in the SFCEVP is followed by a diminution (increment) of the soil moisture. In contrast, positive correlations indicate that the soil moisture is more limiting than energy (soil-moisture-limited regions), and the enhancement of the soil moisture is accompanied by an increase in the SFCEVP.

Although positive SMI-SFCEVP correlations are mandatory for land-atmosphere coupling, the linkage between evapotranspiration and the atmospheric part is also necessary (Dirmeyer, 2011). Therefore, correlations between the SFCEVP and the maximum temperature (Tmax) were also computed, as they are commonly used to identify land-atmospheric coupling (e.g., Diro et al., 2014; Seneviratne et al., 2006). In this case, negative SFCEVP-Tmax correlations show a land-atmospheric coupling, where the increase of temperature is a consequence of the decrease in evapotranspiration. A positive correlation, however, indicates that the increase in temperature leads to more evapotranspiration.

Land-atmospheric interactions may be altered by climate change; therefore, variations in the correlations were also analyzed through their delta-changes. A simple bootstrapping procedure was used to test the significance of the present SMI-SFCEVP correlations as well as the differences with respect to the future, both at the 90% confidence level.

### 2.4. Variables involved in the hydrologic and energy balance

Atmospheric and land variables are closely related through the interchange of water and energy. Therefore, to complete this study, projections in variables involved in both the water and energy balance were also analyzed.

Concerning the water balance, changes in the soil moisture (excluding lateral exchange) are the result of differences between inputs, i.e., total precipitation (prt), and outputs, i.e., evapotranspiration and runoff. Thus, changes in any of the terms lead to modifications in the others. For this reason, variations in the prt and runoff were analyzed to elucidate the potential drivers of soil moisture changes. Additionally, and because land-atmosphere coupling is strongly related to convective precipitation (prc), projections of this component were separately examined. Meanwhile, changes in terms of the energy balance were analyzed through the projections for Tmax, the daily temperature range (DTR), and incoming solar radiation at the surface (SWin). Tmax was selected instead of the daily mean temperature because soil conditions mainly affect the daytime temperature, with the upward fluxes of sensible heat being higher than those at night.

As for soil-related variables, seasonal differences between the future and present periods were used to investigate projected changes, and these were checked according to the non-parametric Wilcoxon-Mann-Whitney Rank Sum test at the 90% confidence level. In this study, the statistical analyses were carried out with MATLAB and the mapping with the Basemap Matplotlib Toolkit from Python.

## 3. Results and discussion

### 3.1. Projected changes in soil moisture and surface evapotranspiration

Fig. 2 displays the seasonally averaged SMI (first and second columns) and the amount of SFCEVP (third and fourth columns) for the WRF simulations driven by the CCSM4 (WRFCCSM) and the MPI-ESM-LR (WRFMPI), for winter (December–February DJF), spring (March–May, MAM), summer (June–August, JJA), and fall (September–November, SON) in

the current period (1980–2014).

The soil water available to plants changes seasonally, with winter/spring (December-May) and summer/fall (June-November) being the wettest and driest seasons, respectively. For the first case, high SMIs were found over the northwestern area where the maximum precipitation occurs (Fig. 2S). Here, values near 1 arise, indicating that the soil at the root-zone reaches its maximum capacity of water to evaporate. In contrast, river valleys such as the Ebro, the Guadalquivir, and the Guadiana show minimum SMIs (around 0.4 and 0.5 for WRFCCSM and WRFMPI, respectively), indicating greater stress conditions for plants. During summer and fall, similar spatial patterns appear, with the SMI generally lower than that in the previous two seasons. For instance, SMIs below 0.1 were reached over river valleys, indicating that the soil water availability is scarce. Regarding differences between the GCM-driven simulations, note that, in general, the WRFMPI provides wetter conditions than the WRFCCSM. However, such an effect is expected to be compensated, at least partly, by the Delta-Change approach. The spatiotemporal patterns of the SMIs in this study (i.e., maximum soil water content in winter-spring and minimum in summer-fall, showing a general northwest-southeast gradient) are similar to those from the soil moisture found by Greve et al. (2013), who evaluated the WRF performance using the ERA-Interim reanalysis as the driving data. This also suggests the suitability of the model performance also when it is driven by the GCMs chosen in this study.

The amount of SFCEVP also varies throughout the year (Fig. 2, third and fourth columns), with fall and winter being the seasons with the lowest amount of SFCEVP. In winter, the spatial variability is the lowest, and the SFCEVP ranges from 0 to 120 mm, approximately. The areas with the highest SFCEVP coincide with those with high SMIs, but SFCEVPs of similar magnitude are also found over the forest in the southwestern and the Guadiana River Basin. Similarly, fall shows low evapotranspiration rates (below 80 mm) for practically all the IP, particularly for the Guadalquivir River Basin. The maximum SFCEVPs, however, appear in the high-altitude regions, forests in Portugal (Fig. 1S), and the northern coastal IP, showing amounts of SFCEVP up to 100 mm.

The highest spatial SFCEVP variability occurs in summertime when a northern-southern gradient is apparent. The maximum SFCEVPs arise over the Cantabrian Coast,

Pyrenees, Central System, Portugal forests, and the Iberian System. Here, high evapotranspiration rates (above 300 mm) are found under an increase in solar radiation in areas where the soil moisture is not limited. In contrast, the southeastern IP together with the Guadalquivir, the Guadiana, and the Ebro Valleys show the lowest SFCEVP (around 20 mm). Note that the radiation in these regions can be even higher than in the previous ones, but as previously mentioned, the soil water here is a limiting factor; therefore, the evapotranspiration is low. However, the spring spatial variability is lower than that from summer. Furthermore, the mean values are generally high (~180 mm) in large parts of the IP due to the concurrence of soil water available to evaporate and the relatively high solar radiation. In this season, minimum SFCEVPs were found over the southeastern IP and in the Pyrenees.

The results of both the soil moisture and SFCEVP also reflect the effects of the vegetation types (Fig. 1S). For instance, during spring, the cropland regions over the Northern Plateau, which have lower canopy resistance to transpiration with respect to their surrounding areas, present higher SFCEVP under similar SMI values. Meanwhile, urban grid-points present anomalous SPCEVP and SMI values (e.g., grid-points corresponding to Madrid, Barcelona, or Porto), suggesting that the WRF has difficulty simulating the land-surface processes in this land-use type, as reported by other authors (García-Valdecasas Ojeda et al., 2020; González-Rojí et al., 2018; Knist et al., 2017).

Regarding projections of the soil conditions, Figs. 3 and 4 display future changes in the seasonal SMI and SFCEVP, respectively, both expressed in relative terms (future *minus* present/present). The stippled areas indicate non-significant changes at the 90% confidence level, and the spatially averaged changes for the whole IP are displayed in the bottom right corners of each panel. In concordance with other studies (Greve et al., 2014; Dirmeyer et al., 2013), our results reveal that increasing GHG concentrations will impact the soil moisture over the IP, particularly under RCP8.5. This increased drying makes the IP a region particularly vulnerable to the desertification process (Dezsi et al., 2018; Gao and Giorgi, 2008).

The most affected seasons by the soil water depletion (Fig. 3) will be those that are the driest in the present (spatially averaged reductions for the whole IP of around 20% and 40% under RCP4.5 and RCP8.5, respectively, in both summer and fall). Concerning the spatial patterns, all simulations present broadly similar behaviors, showing pattern correlations above 0.75 when the RCPs from each GCM-driven simulation are compared, and above 0.6 for simulations between the different GCMs for each RCP. The highest soil dryness occurs in the river basins, particularly over the Guadalquivir and the Guadiana River Basins (diminutions above 55% and 70% under RCP4.5 and RCP8.5, respectively), which are mainly cropland regions. For summer, diminutions of similar magnitude also extend over the Duero and the Ebro Basins.

During winter and spring, however, the changes are more moderate (averaged detriments for the whole IP of up to 9% and 23% under RCP4.5 and RCP8.5, respectively) and even non-significant under RCP4.5 over the northwestern IP. Again, the southern IP (e.g., the Guadalquivir River Basin) presents a clear drying trend. The results also show significant increases (up to 10%) over the Pyrenees, more apparent for the WRFCCSM simulations, and higher under RCP8.5 because of the increased snowmelt (Fig. 3S).

The results of the soil moisture agree in sign with those found by Ruosteenoja et al. (2018) who studied projected changes in the surface soil moisture for the entire European region by using several GCMs from the CMIP5 under RCP4.5 and RCP8.5. They showed that surface-soil moisture is likely to decrease in the IP, with drying consistent throughout all the GCMs analyzed. Moreover, they found that the highest diminutions will be during summer and fall, as our results reveal. Furthermore, for winter and summer, Dirmeyer et al. (2013) recognized a soil-drying trend for the IP using an ensemble of GCMs from the CMIP5 models under RCP8.5. However, the comparison with these studies must be made with caution mainly because, while they used near-surface soil moisture, we analyzed the SMI in the upper 1 m of soil. Furthermore, they used GCM outputs; thus, their spatial resolution was much smaller. In this regard, in a study of the impact of spatial resolution on changes in soil moisture over central-western Europe, van der Linden et al. (2019), found enhanced drying for simulations performed at higher resolution. This latter aspect has an important impact in our region, which is characterized by a strong altitudinal gradient and a high spatiotemporal variability of soil moisture (results not shown).

Although a comparison exercise of our simulations with others performed at regional scale in our study area (e.g., EURO-CORDEX initiative) would allow investigation of the uncertainties associated with the soil drying and its impacts, it is important to consider certain issues in this regard. The simulations performed in this study were configured using an optimal set of parameterizations, selected to simulate the climate over the IP, which is a complex region and is thus more affected by the selection of these parameterizations. In this sense, Jerez et al (2013) empathized that the spread associated with the combination of parameterizations may be of comparable magnitude to those from a multi-ensemble of different simulations performed with different GCMs and RCMs. Furthermore, it is important to consider the difficulty of this comparison because root-zone soil moisture depends strongly on the applied LSM, as well as on other related aspects, such as vegetation type, and soil depth used for simulating the soil conditions (Dirmeyer et al., 2013).

Likewise, all WRF simulations show similar patterns of changes in the amount of SFCEVP (Fig. 4), showing pattern correlations above 0.75, 0.9, 0.85, and 0.55 for winter, spring, summer, and fall, respectively. Comparing these results with the previous ones (Fig. 3), it can be seen that reductions in the SFCEVP are associated with diminutions of the SMI during summer and fall, and over the southernmost area during spring. Summer detriments are notorious, with diminutions over the southernmost IP being around 30% and 50% under RCP4.5 and RCP8.5, respectively. However, increases of the SFCEVP (up to 25%) also appear in this season at high altitude over the northernmost IP (e.g., the Cantabrian Range and the Pyrenees). Similarly, during fall, the SFCEVP undergoes clear reductions. One of the main differences between these two seasons is the increase of evapotranspiration projected over the northernmost IP during summer, not shown for fall. This indicates, in part, the soil water depletion that occurs since spring in this region. Additionally, for this season, while RCP8.5 shows the highest reductions over the southwest (values of around -35%), RCP4.5 indicates maximum decreases over the eastern facade.

During winter, however, extensive regions present non-significant changes, indicating a slight detriment, on average, for the whole IP (reductions of up to 6.5%). Here, all WRF simulations locate the most apparent decreases (up to -40%) over coastal areas in the southern and eastern IP. However, the reductions are located mainly over the eastern façade under RCP4.5, and extended over parts of the central and southern IP for simulations from RCP8.5. Contrariwise, high-altitude regions such as the Pyrenees and the Cantabrian Range present substantial increases in their accumulated winter mean in relation to the present period (above 20% and 55% for RCP4.5 and RCP8.5, respectively). In these latter regions, increases in the soil moisture due to snowmelt (Fig. 3 and Fig. 3S) may lead to more SFCEVP if the temperature increases. Spring shows a similar pattern of changes to winter (spatially averaged reductions of around 3% and up to 10% under RCP4.5 and RCP8.5, respectively). For this season, however, the significant increases are presented over a larger area than the previous one. This together with the results from Fig. 3 suggests that increased SFCEVP would amplify the soil desiccation in these regions. For the Pyrenees, evapotranspiration is even greater than in winter (it rises above 55%) as a result of the occurrence of the temperature increase together with a large amount of snowmelt accumulated since winter.

### 3.2. Land-atmosphere coupling and its future projection

Before analyzing the drivers and potential impacts of soil drying, the model performance of the present-day simulations of the land-atmosphere coupling was first investigated. Fig. 5 displays the seasonal temporal correlations between the SFCEVP and SMI (Fig. 5a), and between the SFCEVP and Tmax (Fig. 5b) in the present-day simulations. The black dots show a non-significant correlation at the 90% confidence level. In general, positive SMI-SFCEVP correlations predominate in all seasons (Fig. 5a), indicating that soil-moisture control is relatively strong for a large part of the IP. This behavior largely agrees with the results from the SFCEVP-Tmax correlations that are mostly negative (Fig. 5b), also revealing a land-atmosphere coupling. However, differences in these correlations appear throughout the year, and along the IP.

During winter, high-altitude regions are controlled by the atmospheric conditions (i.e., significant negative/positive SMI-SFCEVP/SFCEVP-Tmax correlations), except the Pyrenees.

The latter, is at least partly attributable to the fact that these regions are covered by snow (Fig. 3S), resulting in considerably low evapotranspiration rates, and thus non-significant SMI-SFCEVP correlations. However, for a positive temperature anomaly, enhanced snowmelt would lead to an increase in the available soil water, and hence to positive SFCEVP-Tmax correlations as shown in Fig. 5b. For the rest of the peninsula, the soil moisture is the limiting factor, showing significant positive SMI-SFCEVP correlations, in general. However, while the simulation from the WRFCCSM presents land-atmosphere coupling for nearly all the IP (i.e., negative SFCEVP-Tmax correlations), the WRFMPI only shows land-atmosphere coupling throughout the east and southern coastal regions. These discrepancies appear to be associated with the different patterns of large-scale precipitation inherited from the GCMs, presenting drier characteristics in the WRFCCSM compared with the WRFMPI (Fig. 2S).

During spring, more regions present an energy-limited regime (Fig. 5a, MAM). At high altitude, the enhanced snowmelt, together with the precipitation occurring in the preceding months, satisfies the evaporation demand by the atmosphere, which is higher than in winter; thus, these regions remain energy-limited. Meanwhile, the arrival of the frontal systems during winter leads to wetter soil conditions over the rest of the peninsula, shifting the northern IP from a water-limited to an energy-limited region. Concerning land-atmosphere coupling, regions in the south and southeast seem to show negative SFCEVP-Tmax correlations (Fig. 5b, MAM). Again, the results indicate a lower control of the soil moisture for the WRFMPI simulation, according to its wetter character.

For summer, however, the IP generally presents a strong land-atmosphere coupling, as shown by both the SMI-SFCEVP and SFCEVP-Tmax correlations (Fig. 5, JJA). As an exception, the northernmost IP remains energy-limited during summer, as indicated by the significant correlations: positive for SMI-SFCEVP and negative for SFCEVP-Tmax. The summer coupling patterns agree with those found in Lorenz et al. (2012), who studied land-climate coupling over the European climate using correlations between the latent heat flux and near-surface temperature from RCM simulations. Analogous to summer, fall presents large areas with significant positive SMI-SFCEVP correlations (Fig. 5a, SON), being the negative values restricted to small regions over the northernmost (i.e., the Cantabrian Range and the Pyrenees). This is translated to a land-atmosphere coupling over many regions throughout the IP showing significant positive SFCEVP-Tmax correlations (Fig. 5b, SON).

From the soil-drying trend (Fig. 3), changes in the soil regimes are expected. Fig. 6 shows the delta-changes (future *minus* present) in the SMI-SFCEVP correlations. The black dots indicate non-significant changes at the 90% confidence level. In general, changes in land-atmosphere interactions appear to be more apparent for those simulations driven by the WRFMPI, particularly for winter, spring, and summer, and in general, under RCP8.5. In the latter scenario, at high altitude, some regions tend to shift from energy-limited to soil-moisture-limited during winter (i.e., differences in the SMI-SFCEVP correlations above 1). In contrast, a slight decrease in the SMI-SFCEVP correlation will occur over some parts of the Pyrenees. All these results agree with the significant changes that appeared in the SFCEVP-Tmax correlations (Fig. 4S). Greater soil control is also shown during spring, where an increase in the SMI-SFCEVP correlation is found over many regions (e.g., the Northern Plateau and a large part of Portugal). Analogously, the SFCEVP-Tmax correlation decreases substantially in these regions. At high altitude, the changes are more moderate than those in the previous season, except for the Pyrenees, where considerable decreases remain during spring. Therefore, in this latter case, a stronger control of the atmosphere is occurring.

During summer, two different behaviors as a consequence of the soil drying seem to appear. On the one hand, the energy-limited areas over the northernmost IP are reduced, particularly at high altitude due to the increase (decrease) in the SMI-SFCEVP (SFCEVP-Tmax) correlations. On the other hand, for the rest of the Peninsula, the opposite behavior occurs; an increase in the SFCEVP-Tmax indicates a weaker land-atmosphere coupling. The latter is expected for a trend toward a shift from transitional to dry climates. However, the results, in general, are non-significant in many cases. For fall, the results seem to show a slight trend toward a stronger land-atmosphere coupling, particularly in the northernmost areas.

Our findings are partly consistent with those found by Dirmeyer et al. (2013), who studied global trends in the land-atmosphere interactions under RCP8.5, establishing a stronger land-atmosphere coupling during winter over the IP. In the same way, they found stronger coupling in the northern IP, together with a diminution in the south for the summer.

**3.3. Changes in the hydrologic balance**

Changes in the soil moisture depend on variations in evapotranspiration, runoff, and precipitation. Therefore, the projections in the seasonal prt (Fig. 7) and surface runoff (Fig. 8) were analyzed. In general terms, the simulations driven under the two RCPs present similarities in their spatial patterns of precipitation change, showing pattern correlations above 0.5. More differences appear for the surface runoff, especially in summer, when the pattern correlations are below 0.3 in both GCM-driven simulations.

All WRF simulations indicate, on average, slight changes in the winter prt (Fig. 7, DJF). However, reductions are significant only over certain high-altitude regions and in the south-southeastern IP, not in all simulations. Consequently, the surface runoff also decreases in these regions (Fig. 8, DJF). In contrast, certain parts of the IP present slightly increased precipitations, but they are statistically non-significant at the 90% confidence level in nearly all the simulations, and in nearly all the cases. The latter mostly results in an enhanced surface runoff, which is significant especially under RCP8.5 and is shown over the Northern Plateau, where increases above 50% appear. In addition, runoff increases are also found in the Pyrenees, showing changes with respect the historical simulations above 80% in all WRF simulations. Therefore, in these cases, the surface runoff could be considered to prompt soil-drying. Indeed, increases in the surface runoff appear together with an overall reduction in the total runoff (Fig. 5S, DJF). As an exception, the Ebro River Valley and the Pyrenees show significant increases in the total runoff in the simulations under RCP8.5.

Reduction in the prt is already notorious for spring (average variations in the prt are below -20% and -41% under RCP4.5 and RCP8.5, respectively). In this season, the WRF simulations indicate significant changes over the eastern IP (Fig. 7, MAM). The spring prt is still strongly associated with the large-scale circulation, with the non-convective precipitation accounting for, on average, 80% of the prt in our simulations for the whole IP. Hence, the prt reductions are mainly associated with changes in the large-scale patterns as also indicated by the diminution in the non-convective precipitation (result not shown). These changes are more apparent for higher GHG concentrations, showing certain differences in their spatial patterns under the two RCPs (spatial correlation patterns of around 0.6 and 0.9 for WRFCCSM and WRFMPI, respectively). Accordingly, the spring surface runoff (Fig. 8, MAM) generally reduces over regions where the precipitation is substantially declined. The latter, in turn, results in a diminution of water resources, which are more notable when the underground runoff is considered (Fig. 5S).

The highest negative trends in the prt occur in summertime, showing clear decreases under RCP8.5 in nearly all the IP (Fig. 7, JJA). In the latter scenario, diminutions of approximately 50% on average arise in both GCM-driven simulations, reaching values of below -70% in the southernmost regions, extending over the eastern coast. For this season, the synoptic scale is weaker (non-convective precipitation accounts for < 50% on average for the whole IP in the current simulations), with the local effects being more relevant (Jerez et al., 2012). Hence, reductions in the prt are likely to be caused by the diminution in both the convective and non-convective precipitation. In this regard, pronounced decreases in the non-convective precipitation are also projected by the future simulations (results not shown), as previous studies for this region revealed (e.g., Jerez et al., 2012). Summer precipitation reductions translate into significant decreases in both the surface runoff (of around 60% on average under RCP8.5) and the total runoff (Fig. 5S, JJA). Moreover, for this season, the results from the runoff seem to indicate important differences between RCPs, with RCP8.5 showing a clear trend toward decreases in the runoff.

During fall, reductions in the prt are also substantial (Fig. 7, SON), being above 7% under RCP4.5 and around 30% under RCP8.5. For this scenario, the highest decreases appear over the southern-half peninsular, showing values of up to -60%. This indicates that the prt decline is a main driver of drying. Curiously, while the WRFCCSM presents very similar changes to the WRFMPI under RCP8.5, higher discrepancies are found under RCP4.5. Reductions in the surface runoff (Fig. 8, SON) and total runoff (Fig. 5S, SON) also appear where the prt decreases, occurring in a large part of the IP under RCP8.5 (diminutions of approximately 40% and 25%, on average, for the entire IP for the WRFCCSM and WRFMPI, respectively).

The changes found in the precipitation patterns agree with other studies. For instance, Giorgi and Lionello (2008) found reductions in the winter precipitation over the Mediterranean region due to the northward shift of mid-latitude storm tracks. Thus, our results indicate that reductions in the prt are associated primarily with diminutions in large-scale precipitation, which will be reduced in the south and slightly increased in the north. Similarly, Soares et al. (2017) found non-significant increases in the winter precipitation over Portugal using WRF simulations and the EURO-CORDEX multi-model ensembles under the same RCPs used in this investigation. Moreover, both studies recognize reductions in the precipitation during spring, summer, and fall, which will be higher under higher GHG concentrations. In the same way, Argüeso et al. (2012b), using a set of WRF simulations centered over the IP, identified noteworthy reductions in summer precipitations, as well as increases, non-significant in many cases, during winter.

Despite most results showing the same broad trends, certain discrepancies appeared between the two GCM-driven simulations, particularly for RCP4.5 and during fall. Note that climate models commonly suffer from large biases in climate simulations, both in soil conditions (Seneviratne et al., 2006) and general oceanic and atmospheric conditions (Li and Xie, 2013; Li et al., 2017), which affect the future projections. Argüeso et al. (2012a) found that GCMs tend to force the WRF toward excessive zonal circulation and strengthen the north–south pressure gradient over the north Atlantic region, resulting in biases in precipitation over the IP.

In terms of land-atmosphere coupling, the prc is the precipitation component with the most dominant role, so its changes with respect to the present conditions were explored separately and are displayed in Fig. 9. The winter prc is almost negligible and in some regions decreases. However, all simulations show significant increases (above 80%) across the eastern area, which showed a land-atmosphere coupling in both future and present simulations (Fig. 5 and 6). Likewise, fall prc is projected to increase in some areas in the simulations driven by the drier GCM (i.e., WRFCCSM). Specifically, the arrival of the first frontal systems during this season may lead to precipitation recycling. That is, positive anomalies in precipitation lead to an increment in the precipitation via prc (Guo et al., 2006). Indeed, due to the soil moisture is very low in fall, the SFCEVP will be very limited, being very sensitive to variations in the precipitation (Seneviratne et al., 2010). Therefore, although the large scale mainly drives fall and winter precipitation in the present, this feature might change in the future, being the prc more relevant. However, relationships of the causality between the soil moisture and subsequent precipitation are still unclear, these being strongly influenced by the parameterization schemes used to simulate the climate system in a region (Hohenegger et al., 2009).

In contrast, for spring and summer, and particularly under RCP8.5, a general reduction in the prc is found, suggesting that the prc is reduced as a result of the reduction in the large-scale precipitation. Exceptionally, notable increases in the spring prc appear over the Pyrenees, particularly under RCP8.5, which coincide with the enhanced evapotranspiration that occurs in this region (Fig. 4).

**3.4. Changes in energy balance-related variables**

The IP is likely to undergo an increase in Tmax throughout the year (Fig. 10). The spatial patterns of the changes are similar in all seasons, particularly in spring and summer (pattern correlations above 0.85), with the major difference being the magnitude of the changes between the RCPs.

During winter (spatially averaged changes of around 1.25°C and 2.50°C for RCP4.5 and RCP8.5, respectively), the highest temperature rises occur at high altitude, which are more apparent under RCP8.5. In these regions, positive anomalies in the temperature lead to an increased amount of snowmelt; thus, increasing the net shortwave radiation via the decreased albedo (positive snow-albedo feedbacks). Consequently, the maximum temperature also increases (Rangwala and Miller, 2012). Several authors (Giorgi et al., 1997; Rangwala et al., 2013; Xu and Dirmeyer et al., 2012) have identified the snow-albedo feedback as one of the main mechanisms controlling the temperature at high altitude during cold seasons, which has numerous effects on the different components of the climate system. For our study region, López-Moreno et al. (2008) highlighted the impact that could occur in the future due to the temperature rise and the subsequent depletion in the snowpack over the Pyrenees. Moreover, snow-albedo feedback can trigger other indirect effects. That is, increased soil moisture results from snowmelt (e.g., the Pyrenees in Fig. 3) may lead to more runoff (e.g., the Pyrenees in Fig. 8) and SFCEVP (e.g., the Pyrenees in Fig. 4, DJF). The latter partly suppresses the amplified warming, but may also lead to a diminution in the soil water availability. Therefore, the snow-albedo feedback may act as a potential driver of soil-drying, depending on the original soil moisture state and changes in the snow-cover (Xu and Dirmeyer, 2012). Thus, whilst the Pyrenees seems to show a net increase in soil moisture under RCP8.5, other mountain regions present a soil-drying for the future. Meanwhile, the lowest temperature rise appears over the Northern Plateau, where the increase is below 1°C.

Likewise, the spring presents an elevation dependency in its Tmax changes, showing the highest temperature rise over the Pyrenees and the Cantabrian Range. For this season, the snow-cover remains substantial (Fig. 3S, MAM); therefore, the snow-albedo feedbacks may also amplify the temperature rise. However, in this case, decreases in the SMI (Fig. 3, MAM) suggest that the temperature rise via snow-ice albedo feedbacks finally acts as a soil-drying mechanism. Otherwise, lower warming appears over the regions where enhanced evapotranspiration occurs, again indicating the evapotranspiration as a potential driver of soil-drying.

The IP presents maximum warming in summer, with the temperature rise showing the largest differences between the RCPs in its spatially averaged values (changes of around 2°C and above 4.5°C under RCP4.5 and RCP8.5, respectively). The latter indicates that GHG increases will have a greater impact on the temperature during this season. In soil-moisture-limited regions, the SMI is lower than those in the previous seasons (Fig. 3, JJA), and in response to an increase in Tmax, changes in the partitioning of radiative energy occur, increasing the sensible heat flux *vs.* the latent heat flux. Hence, soil moisture-temperature feedbacks are expected that amplify the temperature rise (Jerez et al., 2012). On the other hand, the increase of the temperature together with the reduction in the prt could induce the transformation of energy-limited regions (i.e., the northernmost IP) to soil-moisture-limited under RCP8.5 (Fig. 6, JJA); thus, also increasing the soil moisture control in this region. In contrast, coastal regions present the most moderate Tmax increase, probably due to the moderating effect of the sea (Gómez-Navarro et al., 2010). The temperature rise is also apparent during fall, showing increases in Tmax with respect to the present period up to 4.45°C, on average, for the whole IP. Note that during this season, the soil remains dry (Fig. 3, SON); thus, the interchange of the sensible heat flux between the land and atmosphere is favored rather than the latent heat flux (i.e., evaporation), leading to further warming (positive soil-moisture-temperature feedbacks).

The DTR is also expected to be modified by the soil condition (Dai et al., 1999). All simulations project significant and positive variations in the DTR (Fig. 11) across practically all the IP during spring, summer, and fall. Larger differences with respect to the present period are presented under the highest GHG concentration scenario and for the summer, when increments in the DTR are of around 1.5°C in both simulations. As expected, the DTR patterns changes are associated with changes in soil moisture, as shown by the correlations between the SMI and the DTR, in both the future and present simulations (Fig. 6S). This fact is also evidenced, for instance, through the spring northwest-southeast gradient resulted from the cooling effect of evapotranspiration over the northwest, which does not occur over the south (Fig. 4, MAM). The influence of soil conditions on the DTR was also found in other studies performed over the IP (Jerez et al., 2012), as well as in other regions around the world (Andrys et al., 2017; Expósito et al., 2015).

Decreases in the DTR, however, are also found during winter, particularly in high-altitude regions, as clearly shown in the Pyrenees where the DTR differs from the current period of around -2°C. In this regard, and in agreement with our results, Rangwala and Miller (2012) pointed out that due to the snow-cover depletion, increases in Tmin (result not shown) are possible if increases in the soil moisture (Fig. 3, DJF) and surface humidity also occur. The latter can favor greater diurnal retention of the solar energy at the surface; thus, amplifying the long-wave heating at night. Otherwise, low DTR differences with respect to the present appear over coastal regions, especially in summer, showing even decreases over the Cantabrian coast (Fig. 11, JJA). As indicated by Dai et al. (1999), the diurnal variations in sea breezes partly attenuate the maximum temperature through advection of air mass with different characteristics (i.e., temperature and humidity).

Changes in SWin under unchanged aerosol concentrations are inversely related to changes in the cloud cover. In this regard, all WRF simulations reveal significant decreases in the winter SWin (Fig. 12, DJF), i.e., reductions ranging from -3 to -10 $W/m^2$. For the winter, decreases in SWin resulting from increased cloud cover seem to be partly associated with changes in the prt (Fig. 7, DJF), which is slightly enhanced in many regions where SWin decreases. The highest differences in relation to the present conditions appear at high altitude (decreases of approximately 10 $W/m^2$), where changes in the snowmelt (Fig. 3S, DJF) could lead to an additional increase in the cloud cover via enhanced evapotranspiration (Fig. 4, DJF). Therefore, the hydrological effects of the snow-albedo feedbacks can lead to a reduction in SWin and the subsequent damping effect in the maximum temperature; thus, also reducing the DTR (Fig. 11). Contrariwise, for the rest of the year, and particularly in the summer, SWin is projected to increase. The increased SWin is particularly apparent under RCP8.5 and for the northern IP, where increments of around 30 $W/m^2$ are reached. This could suggest the occurrence of positive soil moisture-radiation feedbacks, as indicated by other recent studies (Ruosteenoja et al., 2018; Vogel et al., 2018). Consequently, further warming is expected, which would affect the Tmax (Fig. 10, JJA). Under RCP4.5, however, the changes are generally non-significant over a large part of the IP, with the WRFCCSM even indicating significant reductions of around 5 $W/m^2$ over the southern IP during fall.

## 4. Conclusions

This work examined projections at high resolution (10 km) over a topographically complex region, the IP. We mainly focused on land-related variables, which have proved to be better represented by RCMs (van der Linden et al., 2019).

The IP is likely to undergo more arid conditions than in the present period in all seasons, greater in magnitude under RCP8.5. This could have implications in the climate variability, depending on changes in the prevailing climate regime (Seneviratne et al., 2010) and along the year. The latter highlights the relevance of considering the projections for all seasons to identify adequately the impact of climate change (Ruosteenoja et al., 2018). Our analysis indicates that, in general, the climate over the IP will be largely controlled by the soil conditions by the end of this century, even over the regions and seasons in which the atmospheric conditions are not presently affected by soil moisture (energy-limited regions).

Table 1 summarizes the main findings in relation to the potential effects/drivers of soil conditions. In this regard, it is important to consider the difficulty in separating these causes and effects (Seneviratne et al., 2010) due to the complex and nonlinear character of the involved processes. The reduction in precipitation is highlighted as a generally predominant driver of soil-drying, particularly during spring, fall, and summer. This mechanism, which seems to be related with changes in large-scale patterns, will be amplified by the lack of local precipitation, particularly for summer and spring. The latter has important implications in vegetation as this season is the most important for vegetation activities. Meanwhile, reductions in runoff (surface and groundwater), strongest during spring and summer, could mean non-recovery during winter. These findings may pose a tremendous challenge for policymakers as they could have significant impacts on the water resources.

This study also suggests the major role of land-atmosphere feedbacks over the IP. We identified different feedbacks potentially influencing the future climate over the IP: the snow-albedo feedback during winter and spring at high altitude alters the soil moisture and subsequently produces different effects through the land-atmosphere interactions (i.e., the snow hydrological effects). The soil moisture-precipitation feedback leads to further drying during spring and summer, and probably favors precipitation recycling during fall and winter. The soil moisture-temperature feedback occurs through changes in the partitioning of surface radiation in the latent and the sensible fluxes, which in turn intensify the warming, particularly over soil-moisture-limited regions during summer and fall. Finally, the soil moisture-radiation feedback results in increased warming, particularly in summer, and with a potential effect in high-altitude regions during winter. However, the cloud effects in a changing climate remain a challenge for the scientific community because a better understanding of how clouds affect the radiative balance is needed, and then, further analysis is required.

**Acknowledgments**

This work was financed by the Spanish Ministry of Economy, Industry and Competitiveness, with additional support from the European Community Funds (FEDER) [CGL2013-48539-R and CGL2017-89836-R], and the 2014-2020 Operational Programme FEDER Andalusia [B-RNM-336-UGR18]. We thank the ALHAMBRA supercomputer infrastructure (https://alhambra.ugr.es) for providing us with computer resources. We thank the anonymous reviewers for their valuable comments that helped to improve this work.

## Figure Captions

**Fig. 1.** (a) WRF model domain: The outer domain corresponds to the EURO-CORDEX region at 0.44° spatial resolution (d01) and the nested domain spanning the IP at 0.088° spatial resolution (d02); (b) Main topographical features in the IP.

**Fig. 2.** Present-day climatology (1980–2014) of the seasonal SMI (first and second columns) and surface evapotranspiration (SFCEVP, third and fourth columns) for the CCSM4- and MPI-driven simulations (WRFCCSM and WRFMPI, respectively).

**Fig. 3.** Projected changes for the seasonal SMI expressed as relative differences (future *minus* present/present). The columns comprise the different GCM-driven simulations under the two RCPs: WRFCCSM RCP4.5, WRFMPI RCP4.5, WRFCCSM RCP8.5, and WRFMPI RCP8.5. The black dots display the non-significant changes at the 90% confidence level. The spatially averaged changes for the whole IP are indicated in the bottom right corners of each panel.

**Fig. 4.** As Fig. 3, but for SFCEVP.

**Fig. 5.** Present-day seasonal correlation ($\rho$) between the SMI and SFCEVP (first and second columns) and between the SFCEVP-Tmax (third and fourth columns) for the two GCM-driven simulations (WRFCCSM and WRFMPI). The non-significant changes at the 90% confidence level are represented by the black dots.

**Fig. 6.** Projected changes ($\Delta\rho$, future *minus* present) in the seasonal correlation between the SMI and SFCEVP. The columns comprise the two GCM-driven simulations under the two RCPs: WRFCCSM RCP4.5, WRFMPI RCP4.5, WRFMPI RCP4.5, and WRFMPI RCP8.5. The stippled areas indicate the non-significant changes at the 90% confidence level.

**Fig. 7.** Future-to-present relative changes of the seasonal precipitation (prt) for the WRFCCSM and the WRFMPI and under the two RCPs (RCP4.5 and RCP8.5). The stippling indicates the non-significant changes at the 90% confidence level. The bottom right corners of each panel show the corresponding averaged values for the whole IP.

**Fig. 8.** Relative projected changes (future *minus* present/present) in seasonal surface runoff. The black dots indicate the non-significant changes at the 90% confidence level. The spatially

averaged changes for the whole IP are indicated in the bottom right corners of each panel.

**Fig. 9.** As Fig. 7, but for the convective precipitation.

**Fig. 10.** Projected changes (future *minus* present) in seasonal maximum temperature (Tmax, °C) for the WRFCCSM and WRFMPI, and for RCP4.5 and RCP8.5. The black dots indicate the non-significant changes at the 90% confidence level. The bottom right corners of each panel indicate the corresponding average values for the whole IP.

**Fig. 11.** As Fig. 10, but for the daily temperature range.

**Fig. 12.** Projected changes (future *minus* present) in solar incoming radiation at the surface (SWin), expressed in $W/m^2$, for both GCM-driven simulations, and for both RCPs (RCP4.5 and RCP8.5). The stippling indicates the non-significant changes at the 90% confidence level. The lower right corners of each panel indicate the corresponding averaged values for the whole IP.

**Table 1.** Summary of potential drivers, adverse impacts, and land-atmosphere impacts implicated in the projected soil future conditions in the IP.

| | DJF | MAM | JJA | SON |
|---|---|---|---|---|
| *Land-atmosphere coupling* | Slightly stronger especially at high-altitude | Stronger especially at high-altitude | Stronger in northernmost IP and slightly weaker for the rest of the IP | Slightly weaker in all the IP |
| *Soil drying mechanisms* | Enhanced evapotranspiration at high-altitude<br>Large-scale precipitation reductions over the south<br>Increase in surface runoff | Enhanced evapotranspiration over the northwest, and especially at high-altitude<br>Non-convective precipitation reductions amplified by a convective precipitation depletion | Enhanced evapotranspiration over energy-limited regions<br>Precipitation reduction amplified by the convective precipitation depletion | Large-scale precipitation reductions |
| *Adverse impacts of soil drying* | Decreases in total runoff (water resources)<br>Increased convective precipitation over the eastern coasts | Decreases in total runoff (water resources)<br>Enhanced radiation at the surface, and amplified temperature warming | Decreases in total runoff (water resources)<br>Enhanced radiation at the surface, and amplified temperature warming | Decreases in total runoff (water resources)<br>Potential increases in convective precipitation |
| *Potential feedbacks* | Snow-albedo feedback<br>Soil moisture-radiation feedback<br>Soil moisture-precipitation feedback<br>Soil moisture-temperature feedback | Snow-albedo feedback<br>Soil moisture-radiation feedback<br>Soil-moisture precipitation feedback<br>Soil moisture-temperature feedback | Soil moisture-radiation feedback<br>Soil-moisture precipitation feedback<br>Soil moisture-temperature feedback | Soil moisture-radiation feedback<br>Soil-moisture precipitation feedback<br>Soil moisture-temperature feedback |

Figure 1

Click here to download Figure: Fig1.pdf

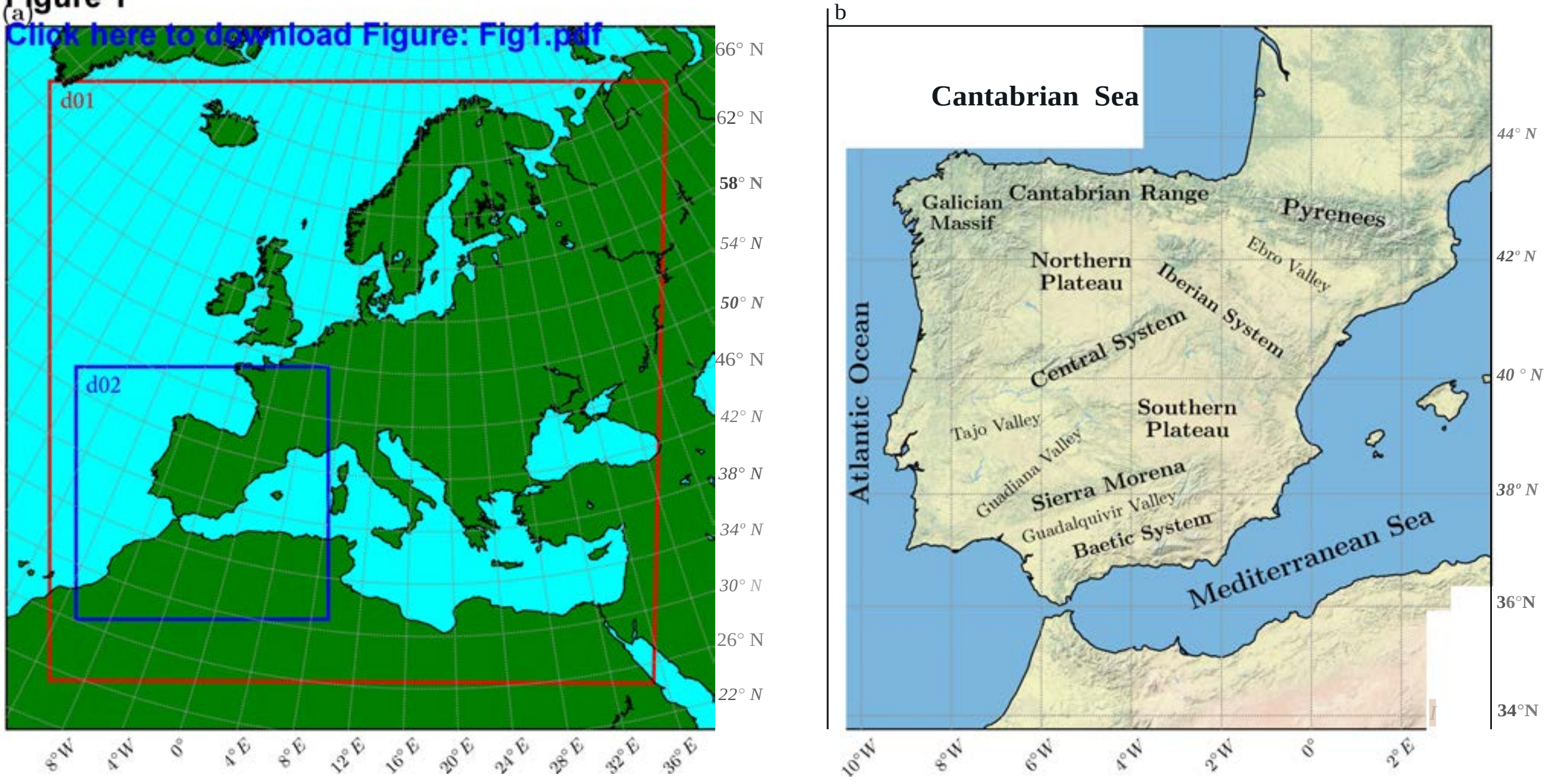

Figure 2
Click here to download Figure: Fig2.pdf

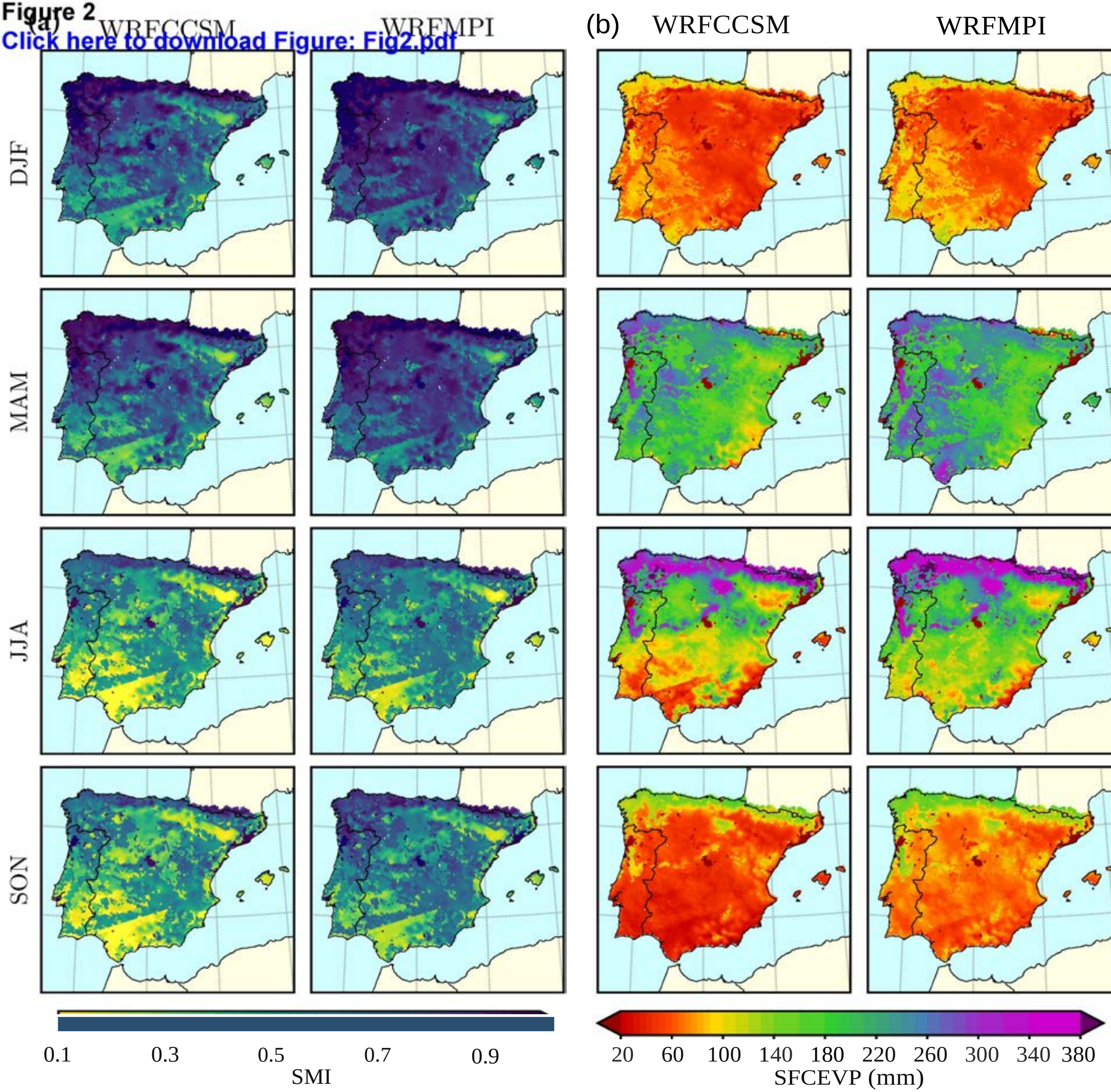

Figure 3

Click here to download Figure: Fig3.pdf

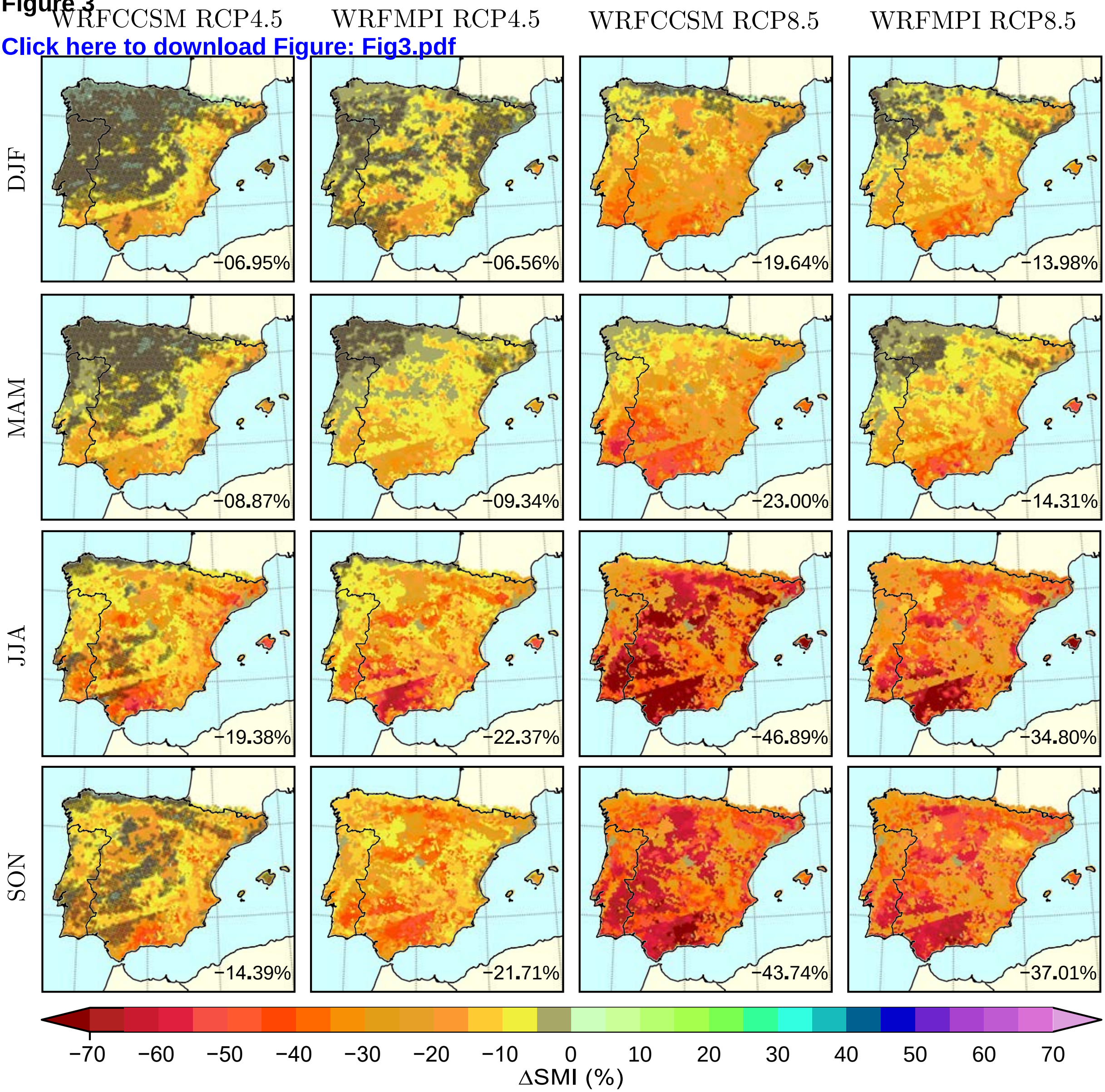

Figure 4

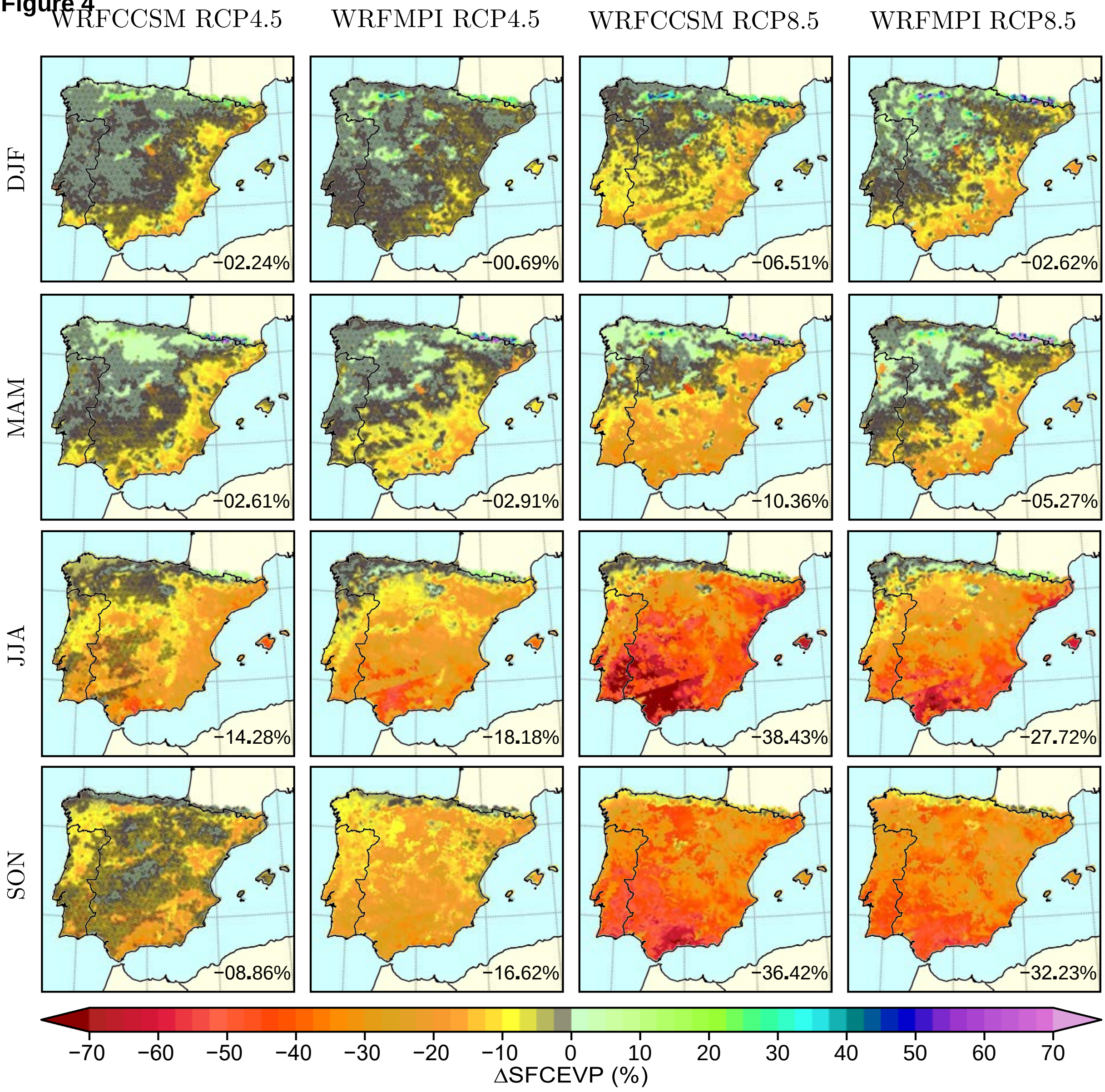
WRFCCSM RCP4.5
WRFMPI RCP4.5
WRFCCSM RCP8.5
WRFMPI RCP8.5
DJF
MAM
JJA
SON
−02.24%
−00.69%
−06.51%
−02.62%
−02.61%
−02.91%
−10.36%
−05.27%
−14.28%
−18.18%
−38.43%
−27.72%
−08.86%
−16.62%
−36.42%
−32.23%
−70 −60 −50 −40 −30 −20 −10 0 10 20 30 40 50 60 70
ΔSFCEVP (%)

Figure 5
Click here to download Figure: Fig5_new.pdf

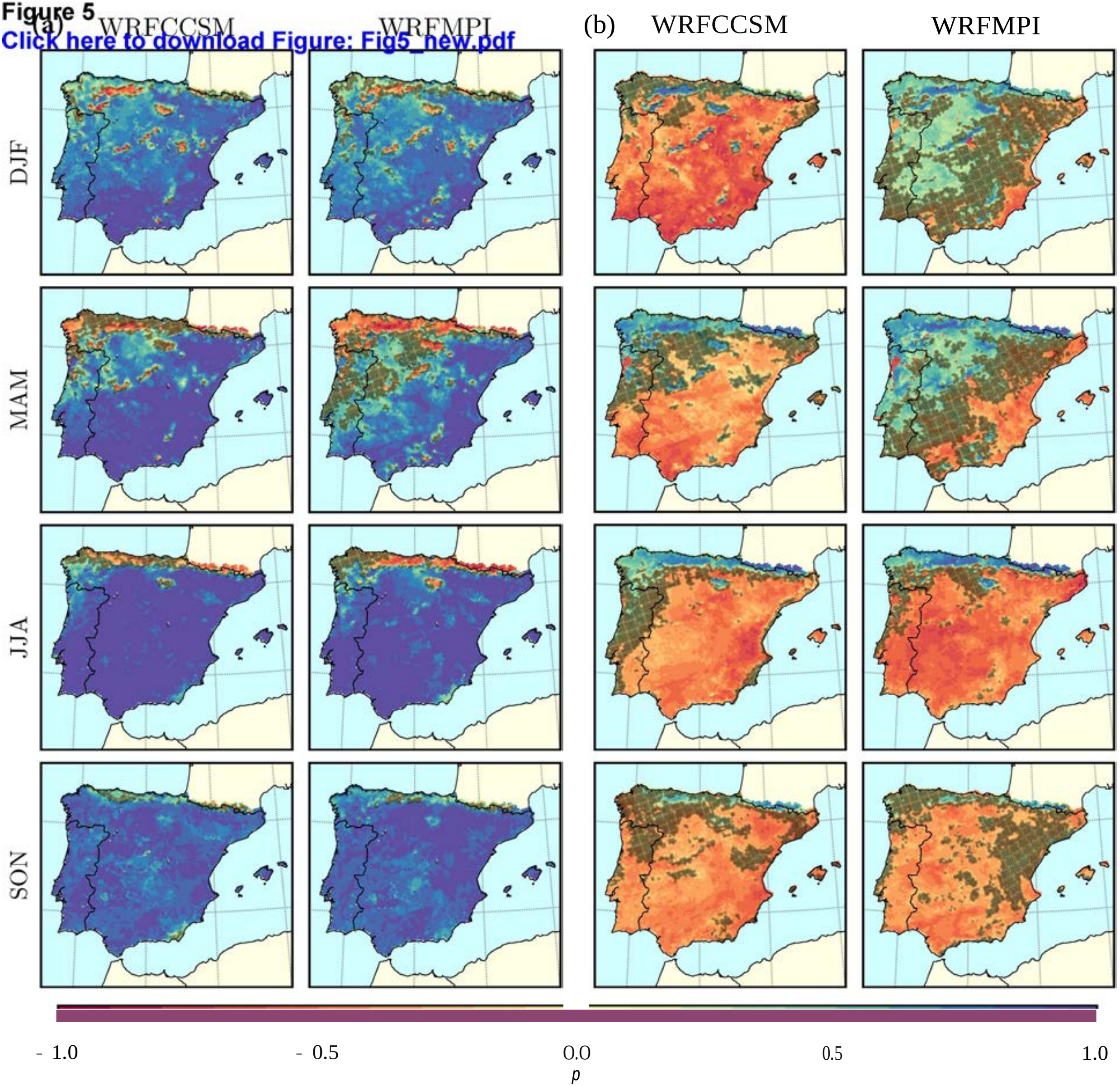

Figure 6
Click here to download Figure: Fig6_new.pdf

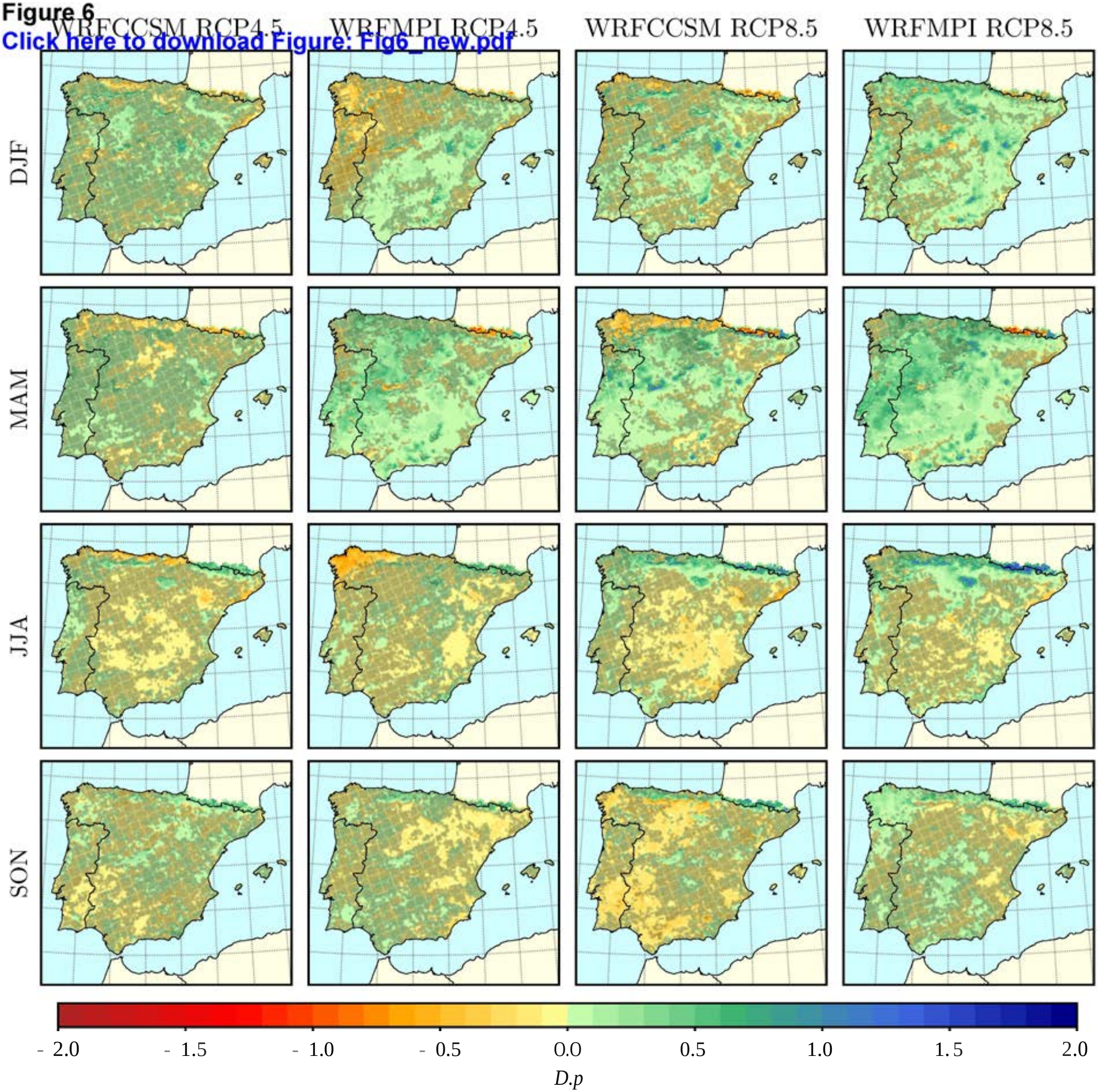

**Figure 7**

Click here to download Figure: Fig7.pdf

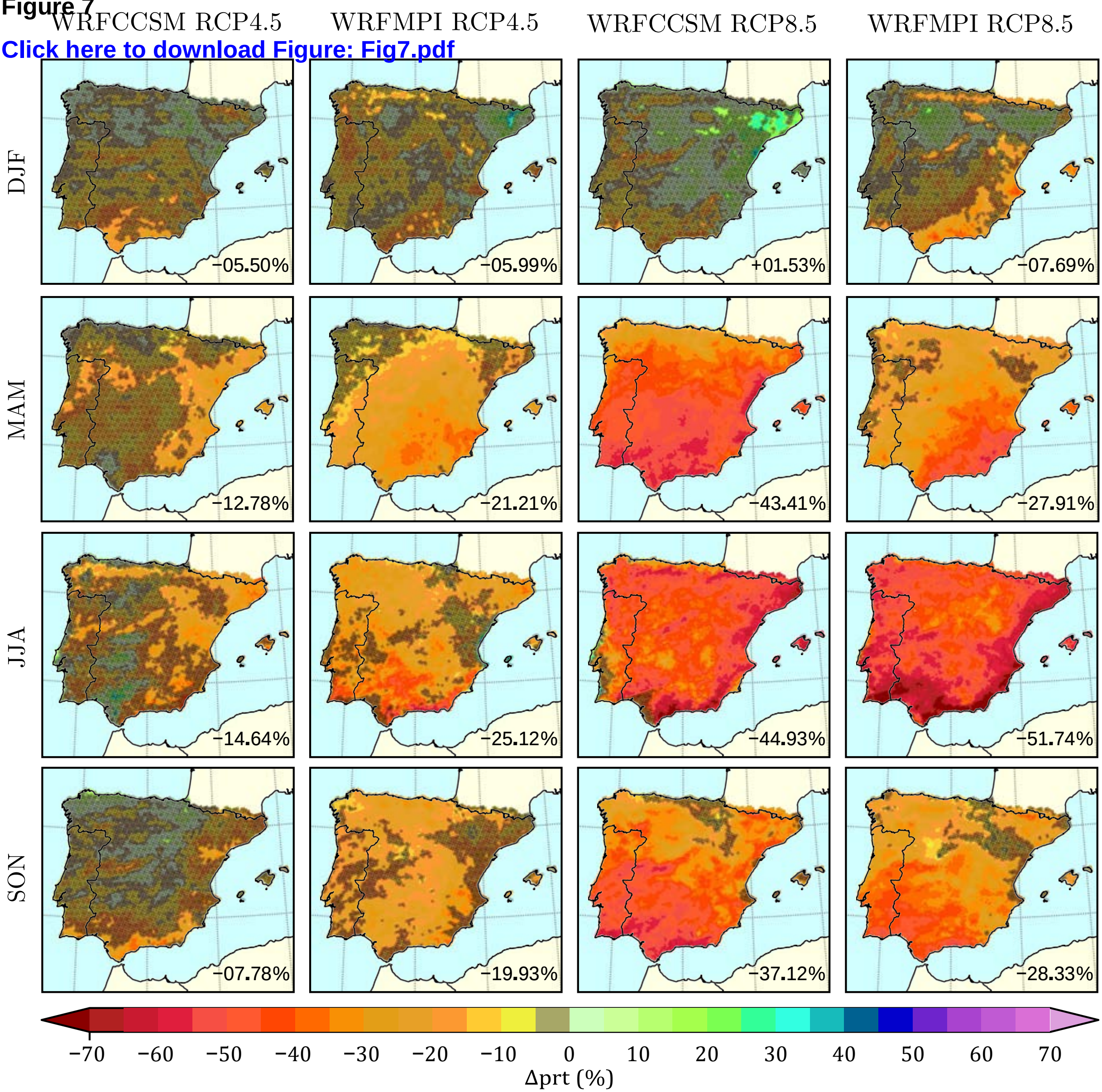

Figure 8

Click here to download Figure: Fig8.pdf

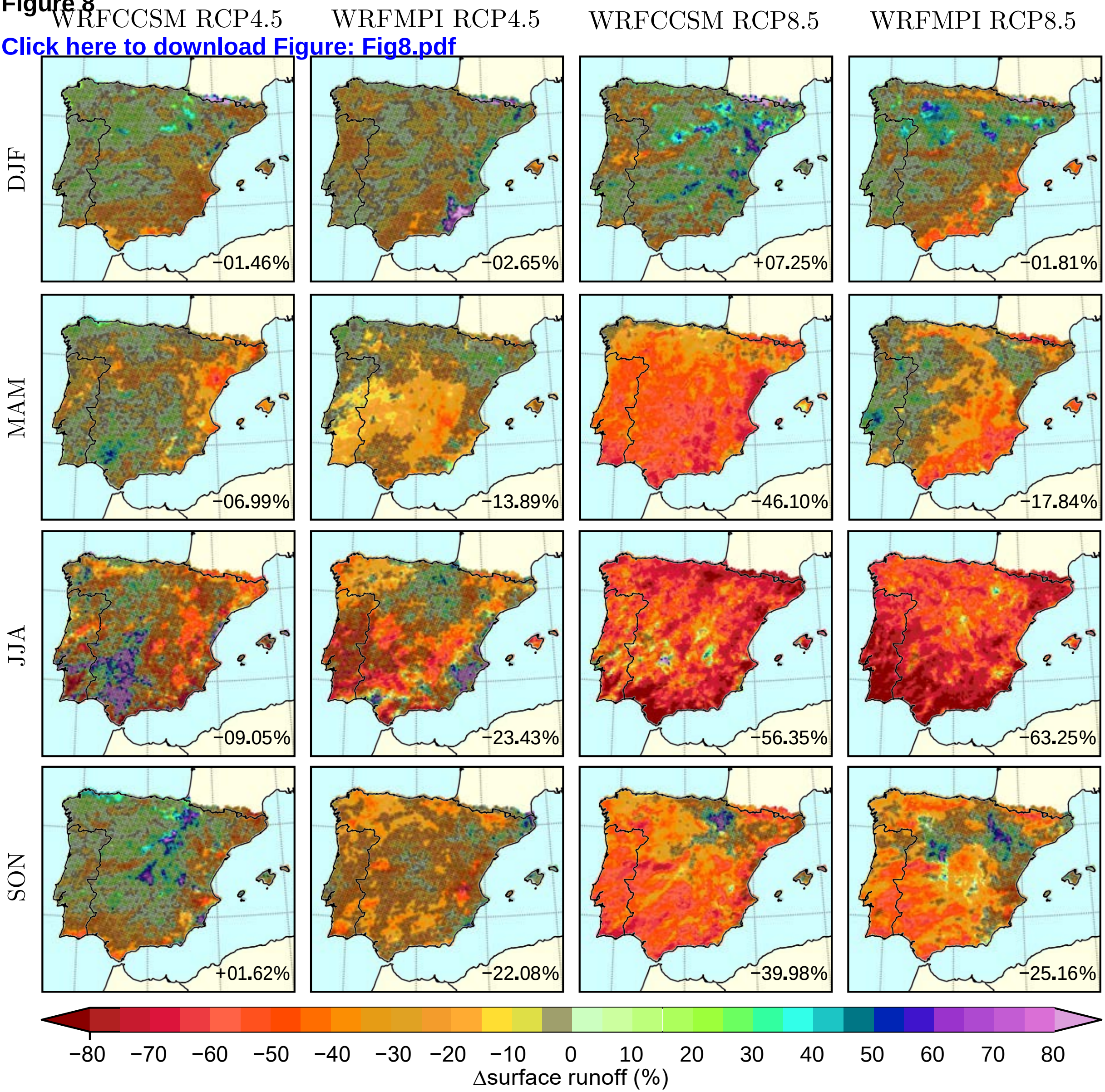

Figure 9

Click here to download Figure: Fig9.pdf

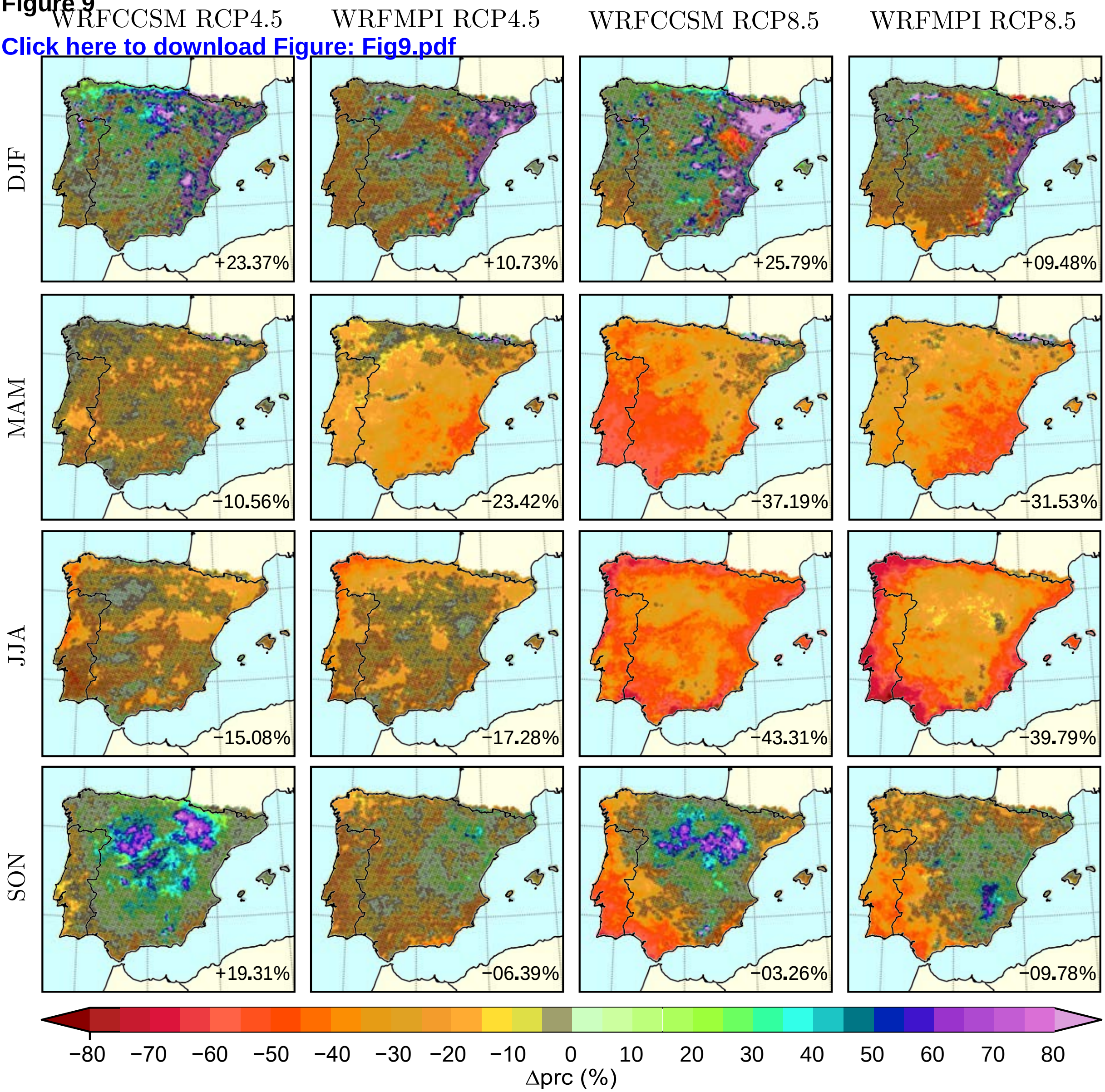

Figure 10

Click here to download Figure: Fig10.pdf

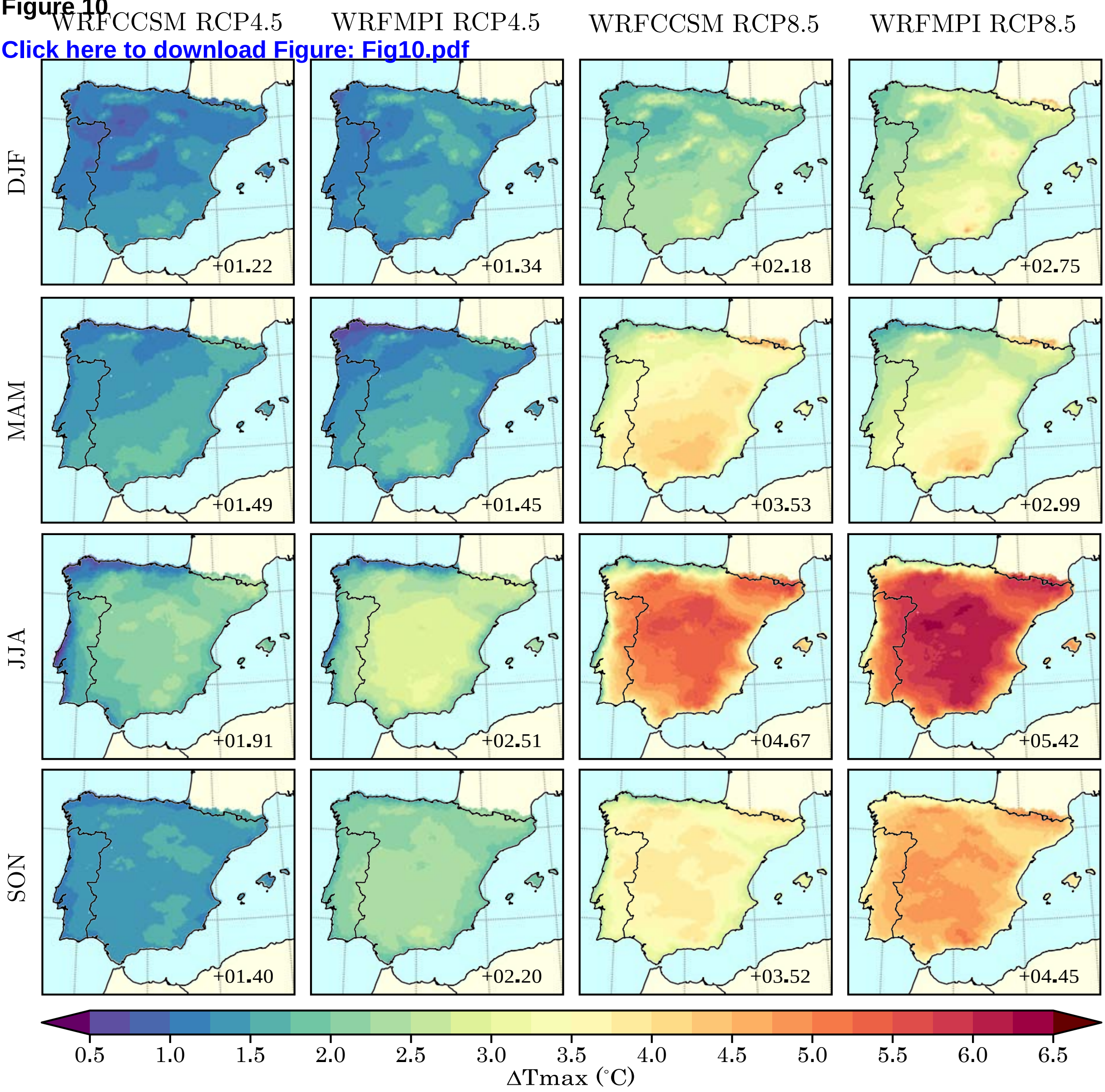

**Figure 11**

Click here to download Figure: Fig11.pdf

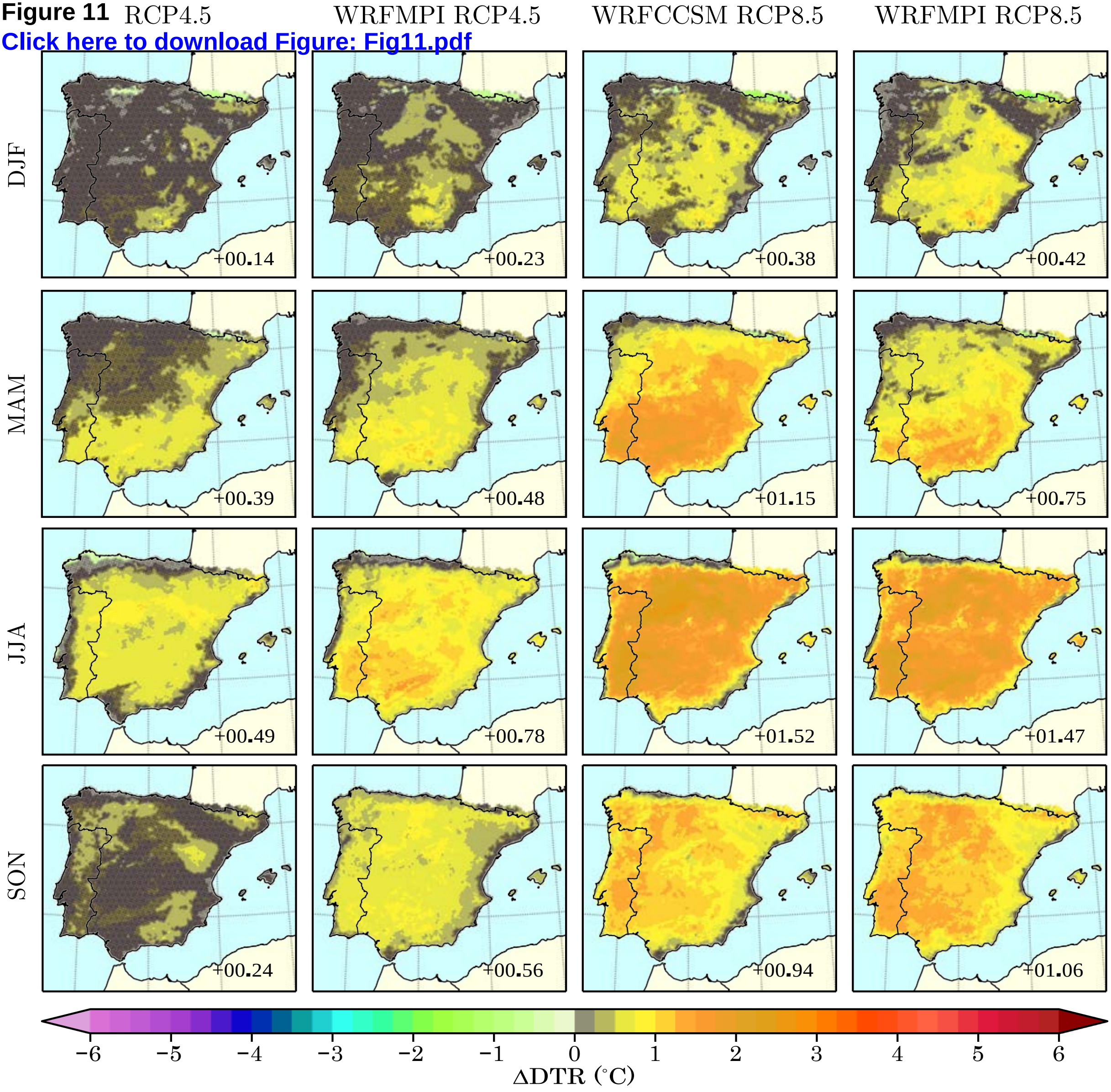

**Figure 12**

**Click here to download Figure: Fig12.pdf**

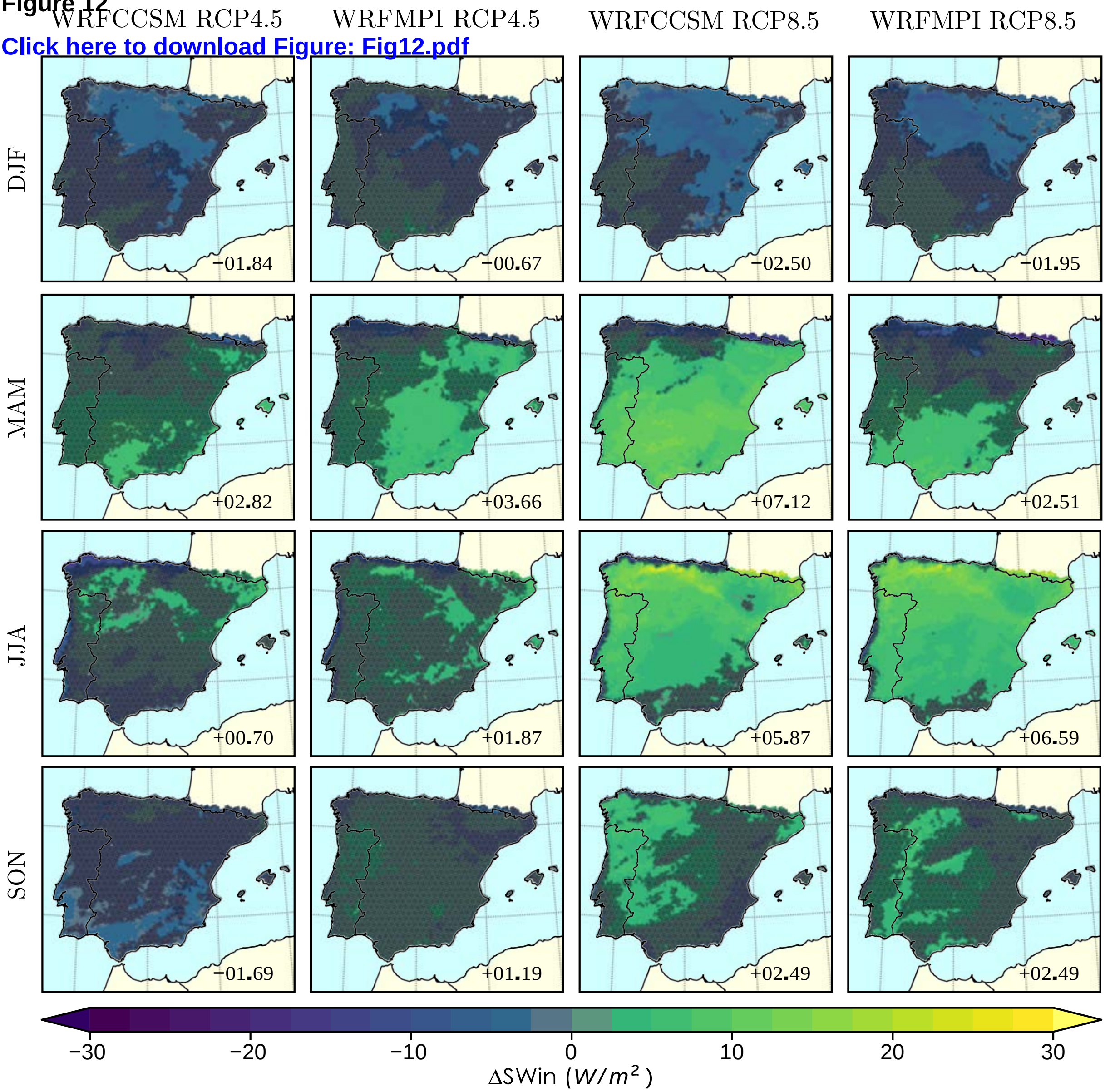

***Credit Author Statement**

**Matilde García-Valdecasas Ojeda**: Conceptualization, Methodology, Software, Validation, Investigation, Data curation, Writing-Original draft preparation.

**Patricio Yeste**: Visualization

**Sonia R. Gámiz-Fortis**: Writing-Reviewing and Editing, Supervision.

**Yolanda Castro-Díez**: Writing-Reviewing and Editing, Supervision.

**María Jesús Esteban-Parra**: Writing-Reviewing and Editing, Supervision, Funding acquisition